\documentclass[aps,pre,reprint]{revtex4-1}
\usepackage[T1]{fontenc}
\usepackage{amsmath}
\usepackage[latin1]{inputenc}
\usepackage{graphicx}

\usepackage{epsfig}

\usepackage{color}

\begin{document}
\title{Phase diagrams of binary mixtures of patchy colloids with distinct numbers of patches: The network fluid regime}

\author{Daniel de las Heras}
\email{delasheras.daniel@gmail.com}
\affiliation{Centro de F\'{i}sica Te\'orica e Computacional da Universidade de Lisboa, Avenida Professor Gama Pinto 2, P-1649-003, Lisbon, Portugal}

\author{Jos\'e Maria Tavares}
\email{josemaria.castrotavares@gmail.com}
\affiliation{Instituto Superior de Engenharia de Lisboa, Rua Conselheiro Em\'{\i}idio Navarro, P-1590-062 Lisbon, Portugal, and Centro de F\'{i}sica Te\'orica e Computacional da Universidade de Lisboa, Avenida Professor Gama Pinto 2, P-1649-003, Lisbon, Portugal}

\author{Margarida M. Telo da Gama}
\email{margarid@cii.fc.ul.pt}
\affiliation{Departamento de F\'{\i}sica, Faculdade de Ci\^encias da Universidade de Lisboa, Campo Grande, P-1749-016, Lisbon, Portugal, and Centro de F\'{i}sica Te\'orica e Computacional da Universidade de Lisboa, Avenida Professor Gama Pinto 2, P-1649-003, Lisbon, Portugal}

\date{\today}
\begin{abstract}
We calculate the network fluid regime and phase diagrams of binary mixtures of patchy colloids, using Wertheim's first order perturbation theory and a generalization of Flory-Stockmayer's theory of polymerization. The colloids are modelled as hard spheres with the same diameter and surface patches of the same type, $A$. The only difference between species is the number of their patches -or functionality-, $f_A^{(1)}$ and $f_A^{(2)}$ (with $f_A^{(2)}>f_A^{(1)}$). We have found that the difference in functionality is the key factor controlling the behaviour of the mixture in the network (percolated) fluid regime. In particular, when $f_A^{(2)}\ge2f_A^{(1)}$ the entropy of bonding drives the phase separation of two network fluids which is absent in other mixtures. This changes drastically the critical properties of the system and drives a change in the topology of the phase diagram (from type I to type V) when $f_A^{(1)}>2$. The difference in functionality also determines the miscibility at high (osmotic) pressures. If $f_A^{(2)}-f_A^{(1)}=1$ the mixture is completely miscible at high pressures, while closed miscibility gaps at pressures above the highest critical pressure of the pure fluids are present if $f_A^{(2)}-f_A^{(1)}>1$. We argue that this phase behaviour is driven by a competition between the entropy of mixing and the entropy of bonding, as the latter dominates in the network fluid regime. 
\end{abstract}

\maketitle

\section{Introduction \label{introduction}}

Recent advances in the engineering of well-defined colloidal particles, with designed surface 
patterning in the nanometer-to-the-micrometer range, open
up the possibility of tailoring their behaviour at the macroscopic level  \cite{vanblaaderen2006,intro1,Granick1,Granick2,Sacanna1,Sacanna2}. 
The resulting anisotropic or patchy particles have become the center of very active research, 
one focus being the development of large-scale fabrication techniques which are required for the exploitation 
of the novel materials in a range of applications \cite{intro2}. 

The current materials revolution based on patchy colloidal particles results from recent breakthroughs in particle 
synthesis and from the fact that these anisotropic colloids may be viewed as the molecules of future materials that can be 
designed to (self)assemble into functional structures. Indeed, the analogy between patchy colloids and molecules provides a 
powerful theoretical framework that exploits the anisotropy of the interactions to control the self-assembly of macroscopic 
structures \cite{intro3}.

Modeling and computational efforts describing patchy particle interactions and their macroscopic properties 
have followed the technological lead and produced a number of interesting results \cite{SciortinoReview2010}. 
The primitive model of patchy colloids consists of hard-spheres with $f$ patches on their surfaces. Patchy particles attract 
each other if and only if two of their patches overlap. The attraction between particles is short ranged and anisotropic: The 
patches act as bonding sites and promote the appearance of  well defined clusters, whose structure and size distribution depend 
on the properties of the patches ($f$ and the energy of attraction) and on the thermodynamic conditions (density and temperature).

\begin{figure}
\epsfig{file=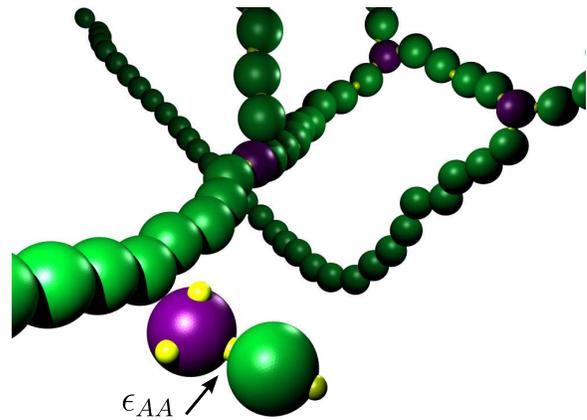,width=3.in,clip=}
\caption{Pictorial representation of a network fluid in a binary mixture of particles with two and tree identical patches. A bond between two bonding sites lowers the energy by $\epsilon_{AA}$.} 
\label{fig1}
\end{figure}

In the last 5 years Sciortino and co-workers investigated, using Monte Carlo simulations, the phase behaviour and the connectivity 
of the fluid phases of the primitive model of patchy colloidal particles, finding a large number of interesting properties. Among others, 
they established that $f$, the number of patches or bonding sites per particle, is the key parameter controlling the location of the 
liquid-vapour critical point \cite{PhysRevLett.97.168301,bianchi2008}. They showed that, for low values of $f$ (approaching 2), the phase 
separation region is drastically reduced and low densities and temperatures can be reached without encountering the phase boundary. These 
low density (``empty'') phases were shown to be network (percolated) liquids (see Fig. \ref{fig1}), suggesting that, on cooling, patchy particles with low functionality 
assemble into glassy states of arbitrary low density (gels). It was also shown that, at the critical point, it is always two network fluids that 
become identical. Very recently, Ruzicka and co-workers have reported the first experimental evidence of empty liquids in dilute suspensions of Laponite \cite{emptyexp}. Their properties were found to be similar to those predicted by the primitive patchy colloidal models.

Remarkably, the results of the simulations of patchy colloidal particles are very well described by classical liquid state theories: 
Wertheim's first order perturbation theory  \cite{wertheim1,*wertheim2,*wertheim3,*wertheim4} predicts correctly the equilibrium thermodynamic 
properties; Flory-Stockmayer \cite{flory1,*stock1,*flory2} theories of polymerization describe quantitatively the size distributions of 
the clusters of patchy particles, including the appearance of network (percolated) fluids. 

Here we extend the investigation of the thermodynamic properties and connectivity of patchy colloidal fluids to binary mixtures of patchy colloidal
particles. As for pure fluids, we will use Wertheim's thermodynamic perturbation theory (in its extension to mixtures) \cite{Chapman:1057} and 
Flory-Stockmayer's theory of percolation (generalized to mixtures - see section \ref{percolationsec}).

In order to quantify the effect of decreasing functionality (approaching 2) on the critical point of pure fluids, Sciortino and co-workers considered 
binary mixtures of particles with 2 and 3 patches \cite{PhysRevLett.97.168301}. However, they (deliberately) ignored the entropy of mixing in the 
theoretical calculations. Instead, they used Wertheim's theory for pure fluids with a non-integer functionality set to the (average) functionality of 
the mixture, assumed to be fixed by the functionality of the mixture at the critical point 
\cite{PhysRevLett.97.168301,bianchi2008}.

A few other studies reported results for the phase diagram of polymer solutions, based on computer simulations and Wertheim-like theories \cite{macdowell:6360,mognetti:044101}. Full critical lines connecting the critical point of the pure polymer to the critical point of the pure solvent 
were reported, and good agreement between the simulation and theoretical results for the coexisting liquid densities 
and compositions was found at low pressures \cite{macdowell:6360}. These models, however, are not directly related to the binary mixtures of patchy 
particles referred above, as they consider monodisperse (fixed length) polymer chains in monomeric isotropic solvents. 

In what follows, we address, explicitly, the interplay between the entropy of mixing and the entropy of bonding in the network fluid regime of models of binary 
mixtures of patchy particles. We restrict our study to mixtures of patchy particles that differ by the number of bonding sites or patches on each species. 

We have found that the phenomenology in the network fluid regime is rich and somewhat surprising as the interplay between the entropy of mixing and the entropy of bonding plays a dominant role, which is absent in other mixtures and/or regimes. In the network fluid regime, the difference between the number of bonding sites is the key parameter controlling the phase equilibria. If one species has at least twice the number of bonding sites of the other, then demixing of two network fluids gives rise to profound changes in the phase diagram of the system. The miscibility at high pressures is also controlled by this difference and closed miscibility gaps are present if the difference between the number of bonding sites is greater than one.

The remainder of the paper is organised as follows. In section \ref{theory} we present the model, 
recall Wertheim's theory for 
binary mixtures (\ref{wertheim}) and generalize Flory-Stockmayer 
theory of percolation to binary mixtures of patchy particles (\ref{percolationsec}); in \ref{particular}, 
we derive expressions for the binary mixtures considered in this paper. In section \ref{Results} we present the results:
phase diagrams (including percolation lines) with emphasis on the network fluid regime and critical properties of several representative mixtures.
Finally, in Sec. \ref{Conclusions} we summarize our conclusions and suggest lines for future research.

\section{Model and Theory\label{theory}}
We consider a binary mixture of $N_1$ and $N_2$ equisized hard spheres (HSs) with diameter $\sigma$. The surface of each species is patterned with a distinct 
number and/or type of bonding sites distributed in such a manner that two particles can form only one single bond, involving two distinct sites one in each particle. A minimum distance between the sites is required to ensure that no sites are shaded by nearby bonds. A pictorial representation of the model under consideration can be found in Fig. \ref{fig1}.

\subsection{Helmholtz free energy: Wertheim's thermodynamic perturbation theory \label{wertheim}}

A detailed description of Wertheim's thermodynamic perturbation theory for pure fluids and fluid mixtures can be found elsewhere \cite{wertheim1,*wertheim2,*wertheim3,*wertheim4,Chapman:1057}. Here we briefly quote the results and set the notation. Within perturbation theory, the Helmholtz 
free energy of a system of particles can be expressed as a sum of contributions from an unperturbed reference system where the particles interact via repulsive forces, here the hard-core repulsions, and a perturbation due to the attractive bonding interactions:

\begin{eqnarray}
f_{H}=F/N=f_{HS}+f_b,
\end{eqnarray}
where $N=N_1+N_2$ is the total number of particles. $f_{HS}$ and $f_b$ are the HS- and bonding- free energy per particle respectively. $f_{HS}$ may be written as the sum of ideal-gas and excess terms: $f_{HS}=f_{id}+f_{ex}$. The ideal-gas free energy is given (exactly) by

\begin{equation}
\beta f_{id}=\ln\eta-1+\sum_{i=1,2}x^{(i)}\ln(x^{(i)} {\cal V}_i),
\end{equation}
with $\beta=kT$ the inverse thermal energy, ${\cal V}_i$ the thermal volume and $x^{(i)}=N_i/N$ the molar fraction of species i. $\eta=\eta_1+\eta_2$ is the total packing fraction ($\eta=v_s\rho$, with $\rho$ the total number density and $v_s=\pi/6\sigma^3$ the volume of a single particle). The excess part, which accounts for the excluded volume of the monomers, is approximated by the Mansoori-Carnahan-Starling-Leland equation of state for HSs mixtures \cite{mansoori:1523}. The latter reduces to the well-known Carnahan-Starling equation of state \cite{carnahan:635} when the species have the same diameter:
\begin{equation}
\beta f_{ex}=\frac{4\eta-3\eta^2}{(1-\eta)^2}
\end{equation} 
The bonding free energy is approximated by Wertheim's thermodynamic first-order perturbation theory \cite{wertheim1,*wertheim2,*wertheim3,*wertheim4}. Let $\Gamma(i)$ be the set of bonding sites or patches on one particle of species $i=1,2$. The bonding free energy per particle is given by \cite{Chapman:1057}
\begin{equation}
\beta f_b=\sum_{i=1,2} x^{(i)}\left[\sum_{\alpha\in \Gamma(i)}\left(\ln X_\alpha^{(i)}-\frac{X_\alpha^{(i)}}{2}\right)+\frac12 n(\Gamma(i))\right],\label{fb}
\end{equation}
where $X_\alpha^{(i)}$ is the probability that a site of type $\alpha$ on a particle of species $i$ is {\it not} bonded and $n(\Gamma(i))$ is the total number of bonding sites per particle of species $i$. The probabilities $\{ X_\alpha^{(i)} \}$ are related to the total density, molar fractions and temperature through the law of mass action:
\begin{equation}
X_\alpha^{(i)}=\left[1+\eta\sum_{j=1,2}x^{(j)}\sum_{\gamma\in \Gamma(j)}X_\gamma^{(j)}\Delta_{\alpha\gamma}^{(ij)}\right]^{-1}.\label{xnotbonded}
\end{equation}
$\Delta_{\alpha\gamma}^{(ij)}$ characterises the bond between a site $\alpha$ on a particle of species $i$ and a site $\gamma$ on a particle of species $j$. For simplicity, we model the interaction between sites by square wells with depths $\epsilon_{\alpha\gamma}$ which depend on the type of bonding sites ($\alpha$ and $\gamma$) but not on the particle species ($i$ and $j$). As a consequence, when the species have the same diameter, the $\Delta_{\alpha\gamma}^{(ij)}$ are independent of the particle species, and can be written as
\begin{equation}
\Delta_{\alpha\gamma}^{(ij)}=\Delta_{\alpha\gamma}=\frac1{v_s}\int_{v_{\alpha\gamma}}g_{HS}({\bf r})\left[\exp(\beta\epsilon_{\alpha\gamma})-1\right]d{\bf r}.\label{delta}
\end{equation}
$g_{HS}({\bf r})$ is the radial distribution function of the reference HS fluid and the integral is calculated over the bond volume $v_{\alpha\gamma}$. We consider that all bonds have the same volume ({\it i.e.}, $v_{\alpha\gamma}=v_b$) and we use the contact value for the radial distribution function. As in previous works \cite{tavares1,*tavares4}, we set $v_b=0.000332285\sigma^3$. Using these approximations Eq. (\ref{delta}) simplifies to
\begin{equation}
\Delta_{\alpha\gamma}=\frac{v_b}{v_s}\left[\exp(\beta\epsilon_{\alpha\gamma})-1\right]A_0(\eta),\label{delta2}
\end{equation}
where
\begin{equation}
A_0(\eta)=\frac{1-\eta/2}{(1-\eta)^3}
\end{equation}
is the contact value of $g_{HS}$. These approximations will affect the results quantitatively but are not expected to change them qualitatively \cite{tavares1,*tavares4}\footnote{We have also studied these mixtures using the ideal-gas approximation for the radial distribution function. The results are qualitatively the same.}. Substituting $\Delta_{\alpha\gamma}$ given by Eq. (\ref{delta2}) into Eq. (\ref{xnotbonded}) we find that $X_\alpha^{(i)}$ depends only on $\alpha$, the type of site ({\it i.e.}, $X_\alpha^{(i)}=X_\alpha,\;\; \forall\; i$) in line with the independent site approximation underlying Wertheim's theory. 

The equilibrium properties of the mixture are obtained by minimising (at a fixed composition $x$, pressure $p$ and temperature $T$) the Gibbs free energy per particle $g=p/\rho+f_H$ with respect to the total density $\rho$ subject to the constraints imposed by the law of mass action. A standard Newton-Raphson method is used to minimise $g$, and the law of mass action is solved simultaneously by a Powell hybrid method or analytically when possible. In what follows we will denote 
the composition of the mixture $x$ by the molar fraction of species $1$: $x\equiv x^{(1)}$ and $x^{(2)}=1-x$. 

Binodal lines are located by a standard common-tangent construction on $g(x)$, which is equivalent to solving the equations for the equality 
of the chemical potentials of both species in the coexisting phases (mechanical and thermal equilibria are satisfied by fixing the pressure and the temperature).

Critical points are computed by determining the states which satisfy the law of mass action and the spinodal condition, $f_{vv}f_{xx}-(f_{xv})^2=0$. In addition, stability requires the vanishing of the third-order derivative in the direction of largest growth \cite{lalmixtures}:
\begin{equation}
f_{vvv}-3f_{xxv}\left(\frac{f_{xv}}{f_{vv}}\right)+3f_{xvv}\left(\frac{f_{xv}}{f_{vv}}\right)^2-f_{vvv}\left(\frac{f_{xv}}{f_{vv}}\right)^3=0,
\end{equation}
where subscripts denote partial derivatives, {\it i.e.}, $f_{xv}$ is the second partial derivative of $f_H$ with respect to the reduced volume per particle $v\equiv 1/\eta$ and the composition $x$ at constant temperature.

The results are presented mostly through temperature-composition diagrams at constant pressure and pressure-temperature projections. The pressure, actually the osmotic pressure in colloidal systems, is difficult to measure in experiments. However, temperature-composition diagrams at constant pressure give a general overview that is somewhat easier to interpret. In addition, we present critical temperature-packing fraction diagrams. From these, one obtains estimates of 
the packing fractions and temperature at two-phase coexistence that may be used to inform experimental studies. We note that the values of the thermodynamic variables at coexistence are sensitive to the model parameters ({\it e.g.}, size and geometry of the molecules, volume of the bonds, interaction bond strength...) and thus serious attempts to predict the behaviour of real systems should be based on more accurate model building. 

\subsection{Percolation in mixtures of particles with distinct numbers or types of bonding sites\label{percolationsec}}
\label{percolation}

Recently \cite{PhysRevE.81.010501} we generalised the Flory-Stockmayer random-bond percolation theory \cite{flory1,*stock1,*flory2} for a model of patchy particles with an arbitrary number of distinct bonding sites (correlated bonding probabilities). The theory, which neglects closed loops, was tested against Monte Carlo simulations and it was found to be quantitatively accurate for systems of particles with three patches of two distinct types \cite{tavares3}. Here,
we extend the theory and derive the percolation threshold for mixtures of patchy particles under the same no-loop assumption.

\begin{figure}
\epsfig{file=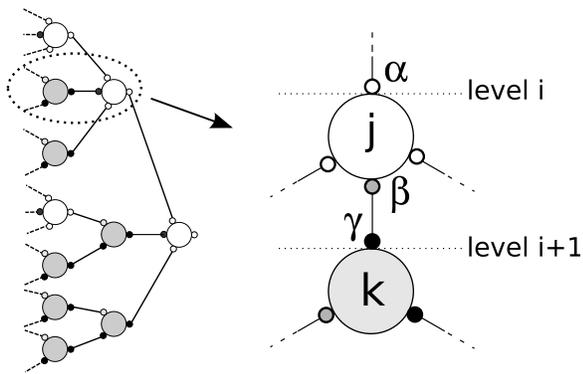,width=3.in,clip=}
\caption{Schematic representation of two bonded particles (right) in a tree-like cluster (left). A particle of species $j$ at level $i$ is bonded to a particle at the previous level $i-1$ through a site $\alpha$ and to a particle at the next level $i+1$ through a site $\beta$. The latter is bonded to a site $\gamma$ on a particle of species $k$.} 
\label{fig2}
\end{figure}

Consider a tree-like cluster and let $n_{i+1,\gamma}^{(k)}$ denote the number of bonded sites $\gamma$ on particles of species $k$ at the level $i+1$. This number is related to the number of all types of bonded sites on both species of particles in the previous level $\{ n_{i,\alpha}^{(j)} \}$ through the recursion relation (see Fig. \ref{fig2} and reference  \cite{PhysRevE.81.010501} for details)
\begin{equation}
n_{i+1,\gamma}^{(k)}=\sum_{j}\sum_{\alpha\in\Gamma_d(j)}\sum_{\beta\in\Gamma_d(j)}p_{\beta_j\rightarrow\gamma_k}\left(f_{\beta}^{(j)}-\delta_{\beta\alpha}\right)n_{i,\alpha}^{(j)}.
\label{levelperco}
\end{equation}
The sum on $j$ runs over the particle species, $j=1,2$ for binary mixtures. $\Gamma_d(j)$ is the set of {\it different} bonding sites on species $j$. $f_{\beta}^{(j)}$ is the number of $\beta$ bonding sites on a particle of species $j$ 
and $p_{\beta_j\rightarrow\gamma_k}$ is the probability of bonding a site $\beta$ on a particle of species $j$ to a site $\gamma$ on a particle of species $k$. Then, the probability of finding a bonded site $\beta$ on a particle of species $j$ is
\begin{equation}
P_{\beta_j}=\sum_k\sum_{\gamma\in\Gamma_d(k)}p_{\beta_j\rightarrow\gamma_k},
\end{equation}
which may be related to the thermodynamic variables through the law of mass action, since 
\begin{equation}
P_{\beta_j}=1-X_\beta^{(j)}.
\end{equation} 
The equations (\ref{levelperco}) are linear in $\{ n_{i,\alpha}^{(j)} \}$ and can be expressed in matrix form
\begin{equation}
\tilde n_i=\tilde T^i\tilde n_0,\label{progressions}
\end{equation}
where $\tilde n_i$ is a vector with components $n_{i,\gamma}^{(k)}$ and $\tilde T$ is a square matrix with entries: 
\begin{equation}
T_{\gamma_k\alpha_j}=\sum_{\beta\in\Gamma_d(j)}p_{\beta_j\rightarrow\gamma_k}\left(f_{\beta}^{(j)}-\delta_{\beta\alpha}\right).
\end{equation}
The matrix $\tilde T$, with dimension equal the number of particle species multiplied by the number of distinct bonding sites, may be diagonalized or 
transformed into Jordan form. In either case the progressions defined by Eq (\ref{progressions}) converge to $0$ if the largest (absolute value) of the eigenvalues $\lambda_{\gamma_k}$ of $\tilde T$ is less than unity, {\it i.e.}, $|\lambda_{\gamma_k}|<1$, $\forall \gamma_k$. Then, percolation occurs when $|\lambda_{\gamma_k}|=1$ for any value of $\gamma_k$.

\subsection{Mixtures of particles with distinct numbers of identical bonding sites\label{particular}}
In the following we focus on the behaviour of mixtures of particles with distinct numbers of identical bonding sites, say $A$. There is then a single bonding energy $\epsilon_{AA}$, which sets the energy scale. We will denote by $f^{(1)}_A-f^{(2)}_A$ a binary mixture of particles of species 1 characterised 
by $f^{(1)}_A$ ($A$) sites and particles of species 2, characterised by $f^{(2)}_A$ ($A$) sites. 

For these mixtures the bonding free energy given by Eq. (\ref{fb}) simplifies to
\begin{equation}
\beta f_b=\langle f \rangle\left(\ln X_A-\frac{X_A}2+\frac12\right),
\end{equation} 
where
\begin{equation}
\langle f \rangle=xf^{(1)}_A+(1-x)f^{(2)}_A,
\end{equation}
is the average functionality or average number of bonding sites per particle at a given composition $x$. The bonding free energy can be split into two terms: 
the bonding energy $U_b$ and an entropic term related to the number of ways of bonding two particles. As the sites are independent, $U_b$ is simply, 
\begin{equation}
\beta\frac{U_b}N=-\frac{1-X_A}2\langle f \rangle\beta\epsilon_{AA},
\end{equation}
while the entropic term is obtained by subtracting the total bonding free energy, given by Wertheim's theory, from $U_b$. 

The law of mass action given by Eqs. (\ref{xnotbonded}) reduces to a single equation for the fraction of unbonded sites,
\begin{equation}
1-X_A=xf^{(1)}_A\eta X_A^2\Delta_{AA}+(1-x)f^{(2)}_A\eta X_A^2\Delta_{AA}\label{xaa-aaa},
\end{equation}
and can be solved analytically. The percolation matrix is a $2\times2$ square matrix with entries
\begin{equation}
\tilde T=
\begin{pmatrix}
p_{A_1\rightarrow A_1}(f^{(1)}_A-1) & p_{A_2\rightarrow A_1}(f^{(2)}_A-1)  \\
p_{A_1\rightarrow A_2}(f^{(1)}_A-1) & p_{A_2\rightarrow A_2}(f^{(2)}_A-1) 
\end{pmatrix}. 
\end{equation}
The probabilities $p_{A_i\rightarrow A_j}$ are found by relating the probability of finding a bonded site $A$, 
\begin{equation}
P_A=p_{A_1\rightarrow A_1}+p_{A_1\rightarrow A_2}=p_{A_2\rightarrow A_1}+p_{A_2\rightarrow A_2},\label{PA}
\end{equation}
to the fraction of unbonded sites $P_A=1-X_A$. A term-by-term analysis of Eqs. (\ref{xaa-aaa}) and (\ref{PA}) gives
\begin{eqnarray}
p_{A_1\rightarrow A_1}=p_{A_2\rightarrow A_1}=xf^{(1)}_A\eta X_A^2\Delta_{AA},\nonumber\\
p_{A_1\rightarrow A_2}=p_{A_2\rightarrow A_2}=(1-x)f^{(2)}_A\eta X_A^2\Delta_{AA}.
\end{eqnarray}
The only non-zero eigenvalue of $\tilde T$ is 
\begin{equation}
\lambda=p_{A_2\rightarrow A_2}(f^{(2)}_A-1)+p_{A_1\rightarrow A_1}(f^{(1)}_A-1),
\end{equation}
and thus the system is percolated if $\lambda>1$. The probability of finding a bonded site at the percolation threshold, $P_p$, can be found by setting $\lambda=1$:
\begin{equation}
P_p=\frac{xf^{(1)}_A+(1-x)f^{(2)}_A}{(1-x)f^{(2)}_A(f^{(2)}_A-1)+xf^{(1)}_A(f^{(1)}_A-1)}.
\end{equation}
Note that this probability depends only on the functionality of the particles and on the composition of the mixture. Furthermore it reduces to the Flory-Stockmayer result 
in the limit of pure fluids: $x^{(i)}=1 \Rightarrow P_p=1/(f^{(i)}_A-1)$.

\section{\label{Results}Results}
We begin by discussing briefly the properties of the pure fluids as they will be used in the following sections in the analysis of the phase behaviour of 
the mixtures. 

\subsection{Pure fluids \label{monocomponent}}

\begin{figure}
\epsfig{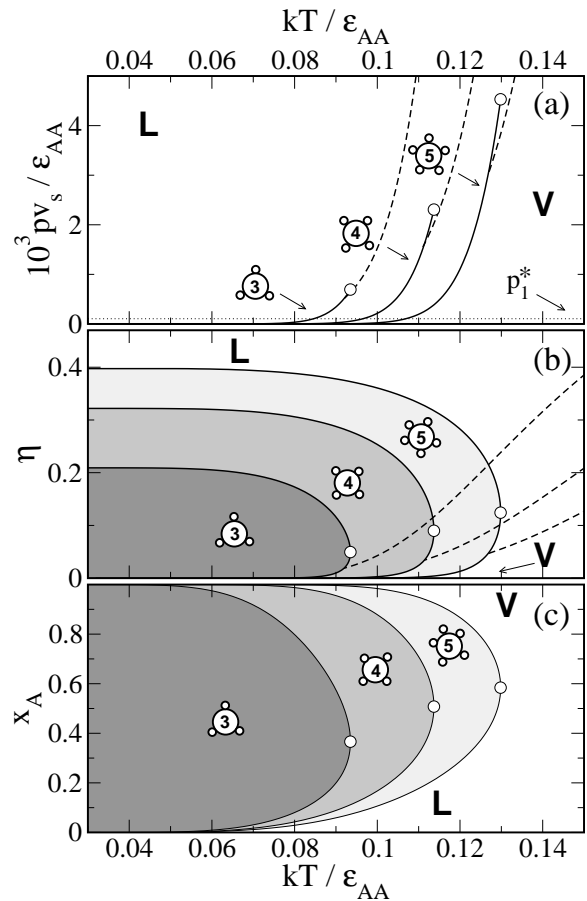}
\caption{Pressure (a), packing fraction (b) and fraction of unbonded sites (c) as a function of temperature in pure fluids of particles with $3$, $4$ and $5$ identical bonding sites. Circles indicate critical points. The dotted horizontal line in (a) marks the pressure $p_1^*=pv_s/\epsilon_{AA}=1.047\times10^{-4}$ referred to in the text. Percolation lines are dashed in panels (a) and (b). The system is percolated above the percolation lines. In panel (c) the percolation thresholds (omitted for clarity) are horizontal lines with values $X_A=1-1/(f_A-1)$, being $f_A$ the number of patches.}
\label{fig3}
\end{figure}

In Fig \ref{fig3} we summarise the properties of pure fluids of particles with $3$, $4$ or $5$ identical bonding sites. These fluids were previously studied using Wertheim's theory and Monte Carlo simulations \cite{bianchi2008}. At temperatures below the critical temperature the system undergoes a first-order phase transition as pressure increases. The transition involves two fluid phases with different densities and fractions of unbonded sites. In what follows we denote these phases by liquid $L$ (higher density and smaller fraction of unbonded sites), and vapour $V$ (lower density and larger fraction of unbonded sites). The $LV$ coexistence line ends at a critical point which moves towards lower pressure and temperature as the number of bonding sites decreases. The packing fraction and the fraction of unbonded sites at the critical point also decrease as the number of bonding sites decreases. As the temperature vanishes, the packing fraction of the coexisting liquid phase saturates to a value that decreases as the functionality decreases. The stability of the liquid phase increases as the number of bonding sites per particle increases but otherwise the phase behaviour is qualitatively the same for all systems. 

The percolation line (dashed lines in Fig. \ref{fig3}) intersects the coexistence line below the critical point on the vapour side. Therefore, the liquid phase is always a network fluid. Near the critical point two percolated states or network fluids coexist, confirming that percolation is a pre-requisite for criticality in systems where the bonding interactions are attractive \cite{prerequisite1,*prerequisite2}.

The behaviour is completely different for a system of particles with $1$ or $2$ bonding sites. In either case there is no $LV$ phase transition. Systems with $1$ or $2$ bonding sites will form dimers and linear chains, respectively. The absence of clustering of the dimers and of branching of the chains prevents the fluids from condensing. A polymerization transition occurs for linear chains in the limit of vanishing temperature, $T\rightarrow 0$ \cite{wheeler:6415}.  

\subsection{Binary mixtures of particles with distinct functionality: $2_A-3_A$ mixture}

We start by considering a $2_A-3_A$ mixture ({\it i.e.}, a binary mixture where species $1$ has two patches of type $A$ and species $2$ has three). The critical properties of this system were investigated recently by grand-canonical Monte Carlo (MC) simulations \cite{PhysRevLett.97.168301}. The results indicate that the critical packing fraction and temperature decrease continuously towards zero as the fraction of $2_A$ particles in the mixture approaches one ({\it i.e.}, the average functionality decreases towards two). The authors of \cite{PhysRevLett.97.168301} mapped the properties of the binary mixture to those of a pure fluid of particles with a (non-integer) number of bonding sites equal to the average functionality $\langle f \rangle$. Although the critical parameters were predicted correctly by this mapping, the effect of the entropy of mixing on the phase behaviour of the mixture was overlooked, and this, together with a full description of the phase diagram, will be addressed in this section.

\begin{figure}
\epsfig{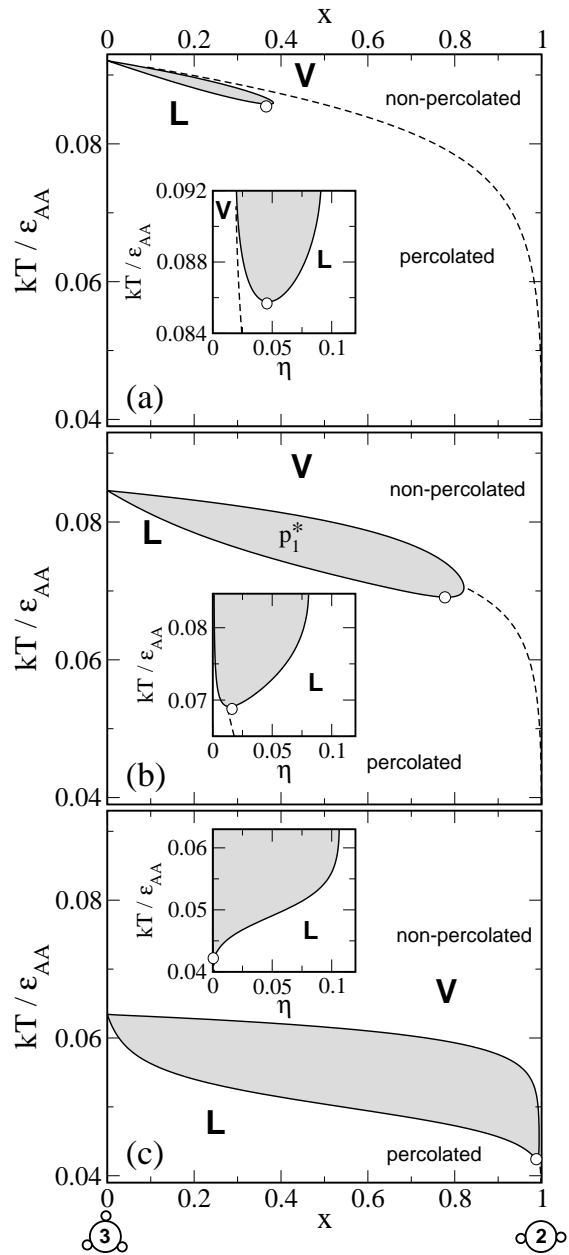}
\caption{Phase diagrams (at different pressures) in the scaled-temperature vs composition ($x=x^{(1)}$) plane of a binary mixture of particles with $2$ (species $1$) and $3$ (species $2$) identical bonding sites. (a) $pv_s/\epsilon_{AA}=5.236\times10^{-4}$. (b) $p_1^*=pv_s/\epsilon_{AA}=1.047\times10^{-4}$. (c) $pv_s/\epsilon_{AA}=1.885\times10^{-7}$. The shaded area indicates the two-phase region. Circles denote critical points. Percolation lines are dashed. The insets depict scaled-temperature vs packing fraction plots along the binodal curve.} \label{fig4}
\end{figure}

In Fig. \ref{fig4} we illustrate the results for the phase diagram of the $2_A-3_A$ mixture in the $T-x$ plane at constant pressure. The diagrams correspond to three distinct pressures below $p_c^{(3)}$, the critical pressure of the pure $3_A$ system. The mixture is always stable for pressures $p>p_c^{(3)}$ (we will return to this point later). At lower pressures the $LV$ phase transition of the pure $3_A$ system shifts from the $x=0$ axis to finite values of the composition, as $2_A$ particles are added to the mixture. The two-phase $LV$ region ends in a lower critical point at $(x_c,T_c)$ and increases as the pressure decreases. The percolation line intersects the binodal on the vapour side, close to the critical point. Below the percolation line the system is a network fluid in the sense that there is a non-zero probability of finding a (more or less transient) infinite cluster. Coexistence involves two network fluids in a finite range of temperatures above the critical temperature, which decreases as the pressure decreases.

\begin{figure}
\epsfig{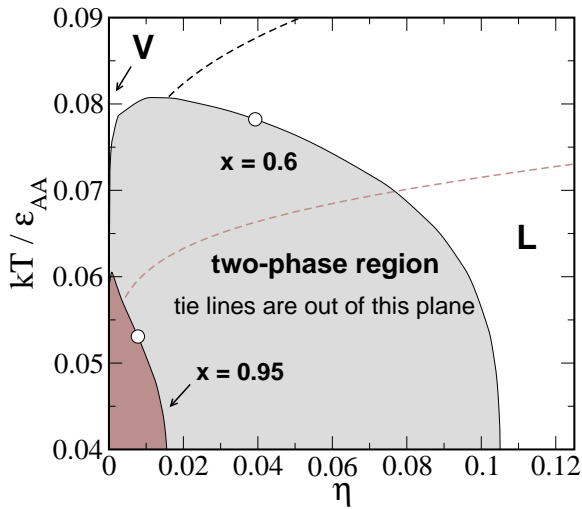}
\caption{Cut of the $(T,\eta,x)$ phase diagram at constant composition for a binary mixture of particles with $2$ (species $1$) and $3$ identical bonding sites. Shaded areas indicate two-phase regions. Open circles denote critical points. Percolation lines are dashed: Black and brown for mixtures with $x=0.60$ and $x=0.95$, respectively.} 
\label{fig5}
\end{figure}

An increase in the fraction of $2_A$ particles reduces the probability of branching, which drives condensation. Therefore, a decrease in the stability of the liquid phase is expected as $x$ increases, and, at a given pressure, the vapour is stable at temperatures below the transition temperature of the pure $3_A$ system (the temperature at the intersection of the binodal with the $x=0$ axis). However, inspection of the insets of Fig. \ref{fig4} ($T-\eta$ plots along the binodal curves) reveals that the density of the liquid phase decreases monotonically with $T$ down to $T_c$. This means that the reduced stability of the liquid phase in the $T-x$ plane is accompanied by an increase in the range of densities where the liquid phase is stable. In order to show this, a cut of the $(T,\eta,x)$ phase diagram at constant composition is depicted in Fig. \ref{fig5}. It is important to note that, in this representation, a tie line connecting two coexisting points in the binodal is out of the plane, except for the critical points and the pure fluid at $x=0$. The two-phase region (shaded area) extends over the whole range of composition (except $x=1$), and decreases as the composition of $2_A$ particles increases. At $x=0.95$, for example, it is possible to stabilise a liquid phase with $\eta\lesssim0.009$ while the lowest density of the liquid in a pure $3_A$ system is $\eta_c^{(3)}\approx0.09$ (see Fig. \ref{fig3}, (b)). In fact, as shown by Sciortino {\it et al.} \cite{PhysRevLett.97.168301}, $\eta_c$ and $T_c$ approach zero asymptotically as the pressure vanishes. 

\begin{figure}
\epsfig{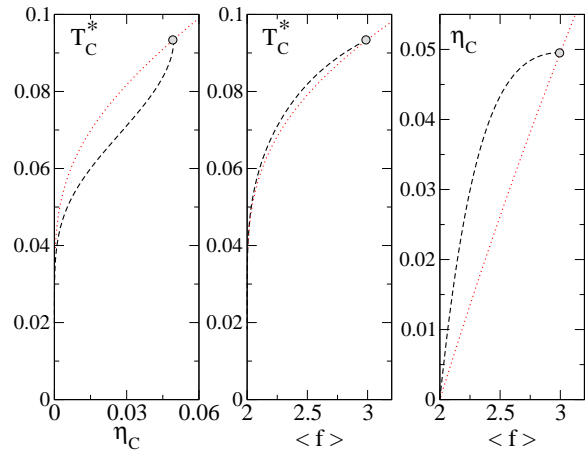}
\caption{Critical properties of a pure fluid with non-integer number of bonding sites (red dotted lines) and a binary mixture of particles with $2$ and $3$ identical bonding sites (black dashed lines). Grey circles mark the points corresponding to a fluid of particles with $3$ identical sites. (left) Scaled-critical temperature $T_c^*=kT/\epsilon_{AA}$ vs critical packing fraction. Middle and right panels depict the scaled-critical temperature and critical packing fraction as a function of $\langle f \rangle$, the average number of bonding sites per particle at the critical point (see text).}
\label{fig6}
\end{figure}

The critical line of the mixture is depicted in Fig. \ref{fig6} where it is compared to the line of critical points of a pure fluid with a number of bonding sites equal to the average functionality of the mixture at the critical point \cite{PhysRevLett.97.168301}. As expected, there are differences between the critical properties of the mixture and those of the pure fluids. The differences result from the composition fluctuations in the mixture, neglected in the mapping to a pure fluid \footnote{Vapour-liquid or liquid-liquid critical points in mixtures are points of incipient material instability rather than mechanical instability as in pure substances (with the exception of azeotropic criticality). See for example \cite{lalmixtures}.}. Nevertheless, the mapping describes correctly the most salient feature of the $2_A-3_A$ mixture critical line: The vanishing of the critical parameters as the composition of the $2_A-3_A$ approaches $x=1$.

Indeed, the agreement between the theoretical predictions for the critical point of pure fluids with average functionality and the MC simulations of the critical line of $2_A-3_A$ mixtures is quite remarkable (see Fig. 3. of reference \cite{PhysRevLett.97.168301}). Nevertheless, the structure of the critical line revealed by the MC simulations signals the effect of composition fluctuations, confirmed by the theoretical calculations of the mixture's critical line (left panel of Fig. \ref{fig6}). In addition, both the MC and the theoretical results for the critical parameters $T_c$ and $\eta_c$ of the binary mixture are above those of the pure fluid with the same functionality, as illustrated in the middle and right panels of Fig. \ref{fig6}. 

\subsection{Binary mixtures of particles with distinct functionalities}
\label{All}

\begin{figure*}
\epsfig{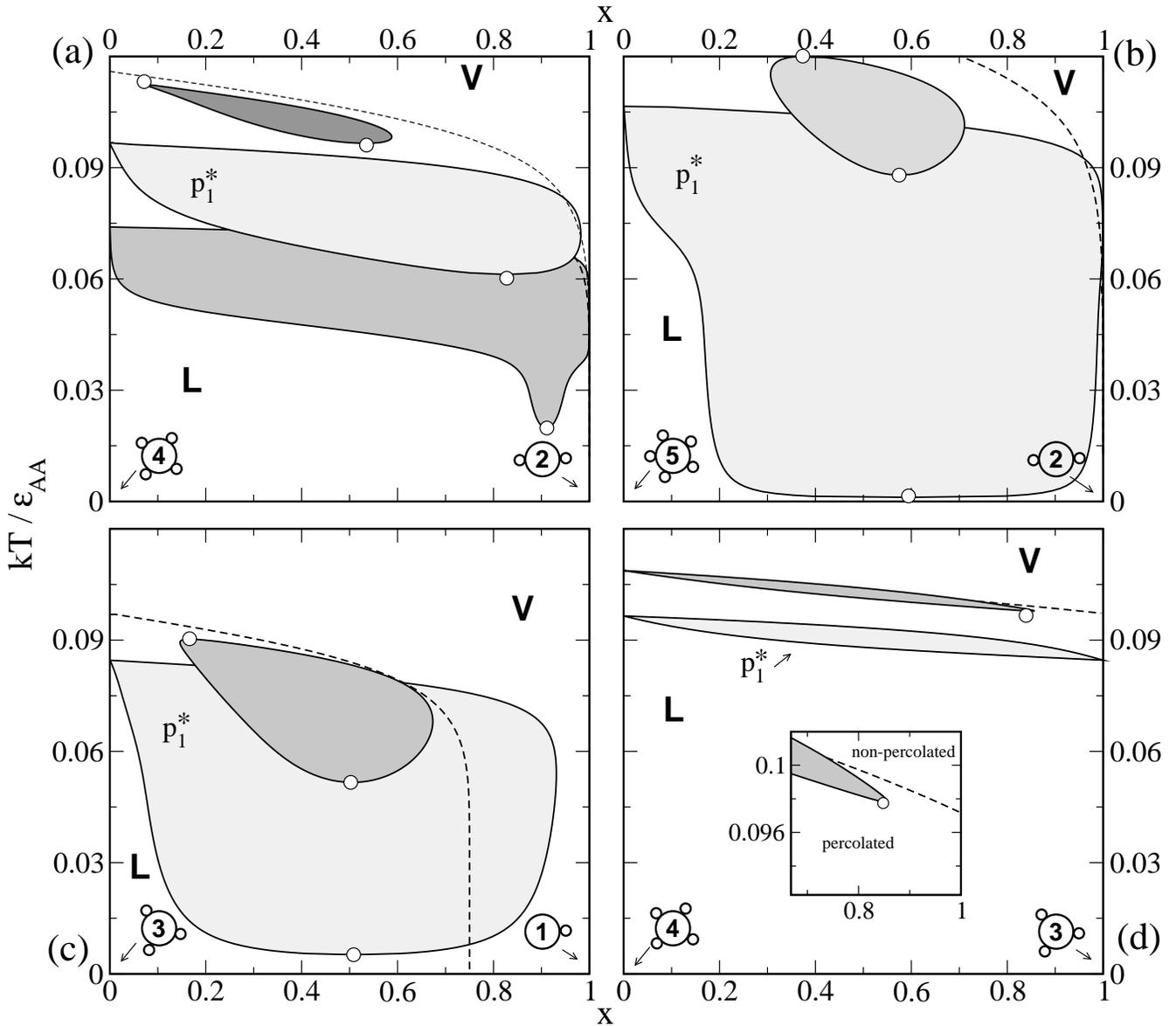}
\caption{\label{fig7}Phase diagrams at different pressures for mixtures of particles with distinct numbers of identical bonding sites in the scaled-temperature $kT/\epsilon_{AA}$ vs composition $x$ plane. Shaded areas are two-phase regions, open circles denote critical points and percolation lines are dashed. (a) $2_A-4_A$ mixture: particles with two A-sites (species 1) and particles with four A-sites (species $2$), pressures: $pv_s/\epsilon_{AA}=2.513\times10^{-3}$ (darkest grey), $p_1^*=pv_s/\epsilon_{AA}=1.047\times10^{-4}$ (lightest grey) and $pv_s/\epsilon_{AA}=1.885\times10^{-7}$ (medium grey). (b) $2_A-5_A$ mixture, pressures: $pv_s/\epsilon_{AA}=1.047\times10^{-2}$ (dark grey), $p=p_1^*$ (light grey). (c) $1_A-3_A$ mixture, pressures: $pv_s/\epsilon_{AA}=1.047\times10^{-3}$ (dark grey), $p=p_1^*$ (light grey). (d) $3_A-4_A$ mixture, pressures: $pv_s/\epsilon_{AA}=1.047\times10^{-3}$ (dark grey), $p=p_1^*$ (light grey) . The inset in (d) is a zoom of the region close to the $LV$ critical point at $p_1^*$.}
\end{figure*}

In this section we consider the binary mixtures: $2_A-4_A$, $2_A-5_A$ and $1_A-3_A$. The results are summarised in the next three figures. Temperature-composition phase diagrams at constant pressure are depicted in Fig. \ref{fig7}. Results are shown at different pressures below and above $p_c^{(3)}$. In all cases the phase diagram at pressure $p_1^*$ is depicted (the phase diagram of a $2_A-3_A$ binary mixture at the same pressure is plotted in panel (b) of Fig. \ref{fig3}). The properties of the critical lines are analysed in Fig. \ref{fig8}, and pressure-temperature projections of the phase diagrams are shown in Fig. \ref{fig9}.

\subsubsection{$2_A-4_A$ mixture: liquid-liquid demixing driven by the entropy of bonding}

In panel (a) of Fig. \ref{fig7} we plot the $T-x$ phase diagram of $2_A-4_A$ mixtures at different pressures below and above $p_c^{(4)}$ (the critical pressure of the $4_A$ fluid). At intermediate pressures (for example $p_1^*$) the $T-x$ phase diagrams are qualitatively the same as those of the $2_A-3_A$ mixture: A two-phase region bounded by a lower critical point starts at the $x=0$ axis and the percolation line intersects the binodal on the vapour side, near the critical point. However, by contrast to the $2_A-3_A$ mixture, the $LV$ density gap along the binodal curve (not shown) has a non-monotonic behaviour. It increases near the $x=0$ axis and then decreases until it vanishes at the critical point. This implies that the drive for phase separation increases when a small fraction of $2_A$ particles is added to the pure $4_A$ fluid, which has not been observed in the $3_A$ case. This behaviour may be understood in terms of the balance between the entropy of mixing and the entropy of bonding. Consider a bond between particles with $2$ and $n$ sites: The resulting structure is a two-particle cluster with $n$ sites available for bonding. The structure that results from a bond between two particles with $n$ sites is a two-particle cluster with $2(n-1)$ sites available for bonding. The latter has $n-2$ additional sites that are available for bonding, and thus the loss in the entropy of bonding increases as $n$ increases, while the gain in the entropy of mixing remains the same. This suggests that the tendency for phase separation increases with $n$, as observed in the $2_A-4_A$ mixture. 

A more striking consequence of a stronger drive for phase separation is the behaviour of the $2_A-4_A$ mixture above the critical pressure of the pure $4_A$ fluid. As shown in panel (b) of Fig. \ref{fig9} there is a range of pressures above $p_c^{(4)}$ where phase separation still occurs. The two-phase region is a closed loop bounded above (below) by an upper (lower) critical point. An example of the $T-x$ phase diagram in this range of pressures is plotted in panel (a) of Fig. \ref{fig7} (darkest grey). The percolation line is always above the phase separation region. As the pressure increases the demixing region decreases, and it vanishes at a given pressure. In panels (a) and (b) of Fig. \ref{fig9} we compare the pressure-temperature projections of the phase diagrams of $2_A-3_A$ and $2_A-4_A$ mixtures. The critical line has a maximum at $p>p_c^{(4)}$ for the $2_A-4_A$ mixture which is absent in the $2_A-3_A$ case. As a result, closed miscibility gaps are found in a range of pressures delimited by this maximum and $p_c^{(4)}$.

Based on the topology of the $pT$ projections, both the $2_A-3_A$ and $2_A-4_A$ mixtures are limiting cases of type I mixtures (according to the classification of 
van Konynenburg and Scott \cite{Scott:49}) where one of the species (particles with $2$ bonding sites) has vanishing critical temperature. In type I mixtures the critical line is continuous and it connects the critical points of the pure fluids. In $2_A-3_A$ and $2_A-4_A$ mixtures the critical line starts at the critical point of species $2$ and tends asymptotically to $T=0$ as the pressure vanishes, $p\rightarrow0$ (species $1$ undergoes a polymerization transition as $T\rightarrow0$). From a topological point of view the significant difference between the critical behaviour of $2_A-3_A$ and $2_A-4_A$ mixtures is that the critical line is monotonic in the former while it is non-monotonic in the latter, with closed miscibility gaps near the critical region of the $4_A$ fluid. However, as we will see now, there is another important difference between both mixtures at very low pressures. 

\begin{figure}
\epsfig{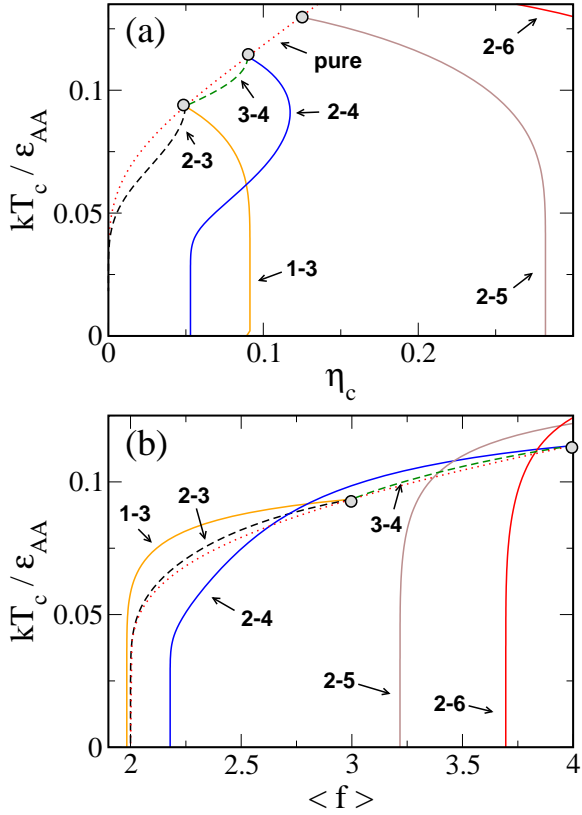}
\caption{Critical properties of binary mixtures of particles with distinct numbers of identical bonding sites. (a) Scaled-critical temperature $kT_c/\epsilon_{AA}$ vs critical packing fraction $\eta_c$. (b) Scaled-critical temperature as a function of the average number of bonding sites at the critical point $\langle f \rangle$. The behaviour of a hypothetical pure fluid with non-integer number of sites is also plotted as a red-dotted line. Grey circles denote critical points of the pure fluids with $3$, $4$ and $5$ sites.} 
\label{fig8}
\end{figure}

Let us focus on the temperature-composition phase diagram of a $2_A-4_A$ mixture at very low pressure (panel (a) of Fig. \ref{fig7} medium gray). We find a large two-phase region, the shape of which suggests the presence (or proximity) of two different phase transitions. Near $x=0$, we observe a $LV$ transition also present in the $2_A-3_A$ mixture, but as the temperature is lowered, a new type of demixing appears (as a bulge) in the phase diagram. In this region two percolated or network fluids coexist (note that this region is well inside the percolated area). We will refer to this as liquid-liquid coexistence in order to distinguish it from the $LV$ coexistence. Both regions form a continuous two-phase region but their different origin is revealed by analysing the critical points (see Fig. \ref{fig8}). As the pressure decreases, the critical temperature vanishes but the critical density tends to a finite value, $\eta_c\rightarrow0.053$. The average functionality at the critical point approaches asymptotically $\langle f \rangle=2.18$ ($x_c=0.911$) rather than $\langle f \rangle=2$ ($x_c=1$). Therefore, although the topology of the mixture is unaffected, the critical line changes its character in a continuous fashion from liquid-vapour (near the critical point of the $4$ patches fluid) to liquid-liquid. The new liquid-liquid transition preempts the liquid-vapour phase transition and there is no liquid-liquid-vapour triple line in this mixture. Nevertheless, $LLV$ triple lines are observed in other mixtures as we will discuss later. 

In Fig. \ref{fig8} we have also presented the critical properties of pure fluids with a number of bonding sites equal to the average functionality of the mixtures at the critical points \cite{PhysRevLett.97.168301}. It is clear that a mapping of the properties of the mixture to those of a pure fluid with the same functionality completely fails in this case, as no empty liquids are found at low pressure. 

\begin{figure}
\epsfig{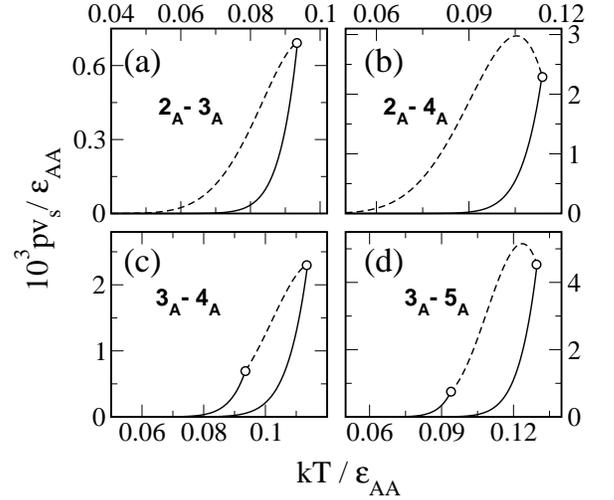}
\caption{Pressure-temperature projections of the phase diagrams of different binary mixtures. (a) $2_A-3_A$ mixture. (b) $2_A-4_A$ mixture. (c) $2_A-5_A$ mixture. (d) $3_A-4_A$ mixture. Solid curves are the liquid-vapour transition lines of the pure fluids. Dashed lines are the critical lines of the mixture. Open circles denote the critical points of the pure fluids. }
\label{fig9}
\end{figure}

Our results strongly suggest that the new liquid-liquid demixing results from a balance between the entropy of bonding and the entropy of mixing. Let us 
consider a bond between two particles with $f^{(2)}_A$ sites. The resulting structure is a two-particle cluster with $2(f^{(2)}_A-1)$ unbonded sites. This corresponds to an increase of $f^{(2)}_A-f^{(1)}_A$ unbonded sites when compared to the structure that results from bonding particles of species 
$1$ and $2$. When this increase is large compared to the number of bonding sites of species $1$, the entropy of bonding gained by bonding particles of species $2$ 
compensates the entropy of mixing lost and drives the liquid-liquid phase separation. Considering only an integer number of bonding sites the condition can be written as:
\begin{equation}
f^{(2)}_A\ge2f^{(1)}_A.
\label{double}
\end{equation}
That is, if one species has at least twice the number of bonding sites of the other, then liquid-liquid demixing, driven by the entropy of bonding, occurs. Nevertheless, if we allow the particles to have a non-integer number of bonding sites, then the condition for finding liquid-liquid demixing is not exactly given by Eq. (\ref{double}). For example, in mixtures where species $1$ has two sites, we have found empty liquids -- and no liquid-liquid demixing -- if $f^{(2)}_A\lesssim 3.99$, while liquid-liquid demixing occurs when $f^{(2)}_A\gtrsim3.99$. \footnote{The value at which the liquid-liquid phase transition occurs is sensitive to the approximation of the radial distribution function $g_{HS}$. For example, using the ideal-gas approximation, we found liquid-liquid demixing when the number of patches of species $2$ is greater than twice the number of patches of species $1$ (rather than at least twice).}. 

We have checked the heuristic argument of Eq. (\ref{double}) by calculating the phase diagrams of other mixtures satisfying this condition. 

\subsubsection{$2_A-5_A$ mixture}

Phase diagrams of the $2_A-5_A$ mixture are depicted in panel (b) of Fig. \ref{fig7} at two different pressures below (light grey) and above (dark grey) $p_c^{(5)}$, the critical pressure of the pure fluid with 5 bonding sites. As in the previous mixture, there is a region of pressures above $p_c^{(5)}$ where closed miscibility gaps are present. In this high pressure region, the phase coexistence involves two percolated fluids. As the pressure decreases, the two-phase region grows very rapidly, occupying a large fraction of the phase diagram (compare, for example, the phase diagram of this mixture at pressure $p_1^*$ with other mixtures at the same pressure).

Topologically this mixture is still a limiting case of type I. The analysis of the percolation threshold and the critical properties (see Fig. \ref{fig8}) indicates the coexistence of two network fluids in the limit of low pressures. As the pressure decreases, the critical temperature tends asymptotically to 
$T_c=0$, but the critical packing fraction and critical composition tend to finite values ($\eta_c\rightarrow0.282$ and $\langle f \rangle\rightarrow3.22$). There is a slight difference in the behaviour of $2_A-4_A$ and $2_A-5_A$ mixtures. In the latter, the critical packing fraction is always higher than the critical packing fraction of the pure fluid with $5$ patches, $\eta_c^{(5)}$, and increases monotonically with decreasing pressure (see panel (a) 
of Fig. \ref{fig8}). In the former ($2_A-4_A$ mixture), the critical packing fraction increases from $\eta_c^{(3)}$, reaches a maximum, then decreases slowly, and tends to a plateau that is lower than the critical density of the pure fluid with $3$ patches.

\subsubsection{$1_A-3_A$ mixture}

We proceed by considering the $1_A-3_A$ mixture ($T-x$ phase diagrams are plotted in panel (c) of Fig. \ref{fig7}) which also satisfies Eq. (\ref{double}). The addition of $1_A$ particles to the $3_A$ fluid drastically reduces both the probability of branching and the average cluster size. At $x\approx0.75$ the predominant structures are isolated clusters of three $1_A$ particles that saturate the bonds of the $3_A$ particles. The most obvious consequence of this is that when $T\rightarrow0$, the percolation line tends asymptotically to $x=0.75$ rather than to $x=1$. Thus, it is not possible to observe a percolated fluid 
if the composition of the mixture is greater than $x=0.75$.

We also found closed miscibility gaps at $p>p_c^{(3)}$; an example is the $T-x$ phase diagram at $pv_s/\epsilon_{AA}=1.047\times10^{-3}$ depicted in panel (c) of Fig. \ref{fig7} (dark grey). The line of critical points, represented in Fig. \ref{fig8}, behaves qualitatively as that of the $2_A-5_A$ mixture and, surprisingly, it is possible to find coexistence between fluid phases with an average functionality smaller than two (note that $\langle f \rangle\rightarrow1.98$ when $T_c\rightarrow0$). Topologically, this is still a type I mixture.

\begin{figure}
\epsfig{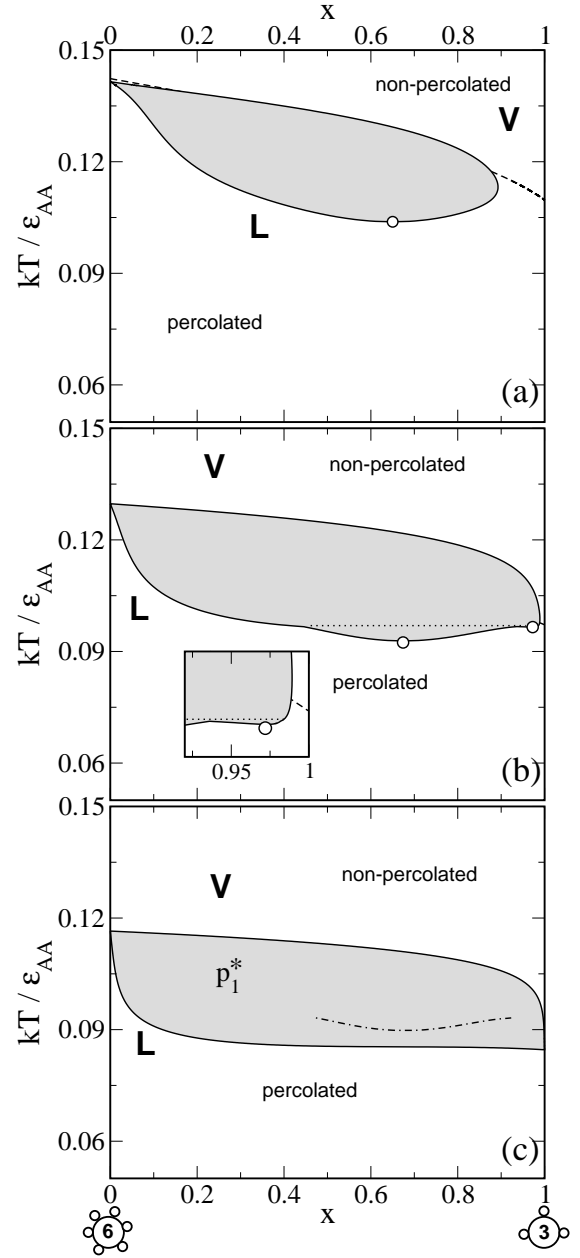}
\caption{Phase diagrams (at different pressures) in the scaled-temperature vs composition ($x=x^{(1)}$) plane of a binary mixture of particles with $3$ (species $1$) and $6$ (species $2$) identical bonding sites ($3_A-6_A$ mixture). (a) $pv_s/\epsilon_{AA}=5.236\cdot10^{-3}$. (b) $pv_s/\epsilon_{AA}=1.047\cdot10^{-3}$. (c) $p_1^*=pv_s/\epsilon_{AA}=1.047\cdot10^{-4}$. The shaded area is the two-phase region. Circles denote the critical points. Percolation lines are dashed. The horizontal dotted line in (b) indicates the position of the triple point. The inset in (b) is a zoom of the region near the $x=1$ axis. In this zoom the percolation line connects the $x=1$ axis to the binodal. The dashed-dotted line inside the demixing region in (c) is the metastable binodal for $LL$ coexistence.} 
\label{fig10}
\end{figure}

\subsection{From type I to type V mixtures: $3_A-f_A^{(2)}$ mixtures}
\label{AA-f2A}

Finally, we focus on $3_A-f^{(2)}_A$ mixtures, in particular, $3_A-4_A$ and $3_A-6_A$. The latter satisfies the liquid-liquid demixing condition, Eq. (\ref{double}), while the former does not. By contrast to the mixtures considered previously, both pure fluids exhibit $LV$ transitions.

The phase diagram of the $3_A-4_A$ mixture is plotted in panel (d) of Fig. \ref{fig7}. At intermediate pressures, between $p_c^{(3)}$ and $p_c^{(4)}$, the phase diagram consists of a small two-phase region bounded by a lower critical point. The percolation line starts at $x=1$ at finite temperature and intersects the binodal on the vapour side, close to the critical point. At pressures below $p_c^{(3)}$ ({\it e.g.}, $p_1^*$) the critical point disappears and there is $LV$ coexistence over the whole range of composition. The liquid phase is always percolated in this range of pressures. Above $p_c^{(4)}$ the system is completely miscible. The line of critical points in a $pT$ projection, represented in panel (c) of Fig. \ref{fig9}, connects the critical points of the two pure fluids and thus it is a standard type I mixture. The temperature and the density on the critical line are similar to those obtained from the mapping to a pure fluid with the same functionality (see Fig. \ref{fig8}). In $3_A-5_A$ mixtures, the only difference is the existence of closed miscibility gaps at pressures higher than $p_c^{(5)}$ (the critical line is non-monotonic. See, for example, the $pT$ projection of this mixture in panel (d) of Fig. \ref{fig9}).

However, new phenomenology is found in the phase behaviour of $3_A-6_A$ mixtures. Three representative phase diagrams are plotted in Fig. \ref{fig10}. At pressures slightly below $p_c^{(6)}$ there is a large two-phase region starting at $x=0$ and ending in a lower critical point at finite values of the composition (panel (a) of Fig. \ref{fig10}). As $p_c^{(3)}$ is approached, a phase transition involving two liquid-like states is clearly visible. There are two critical points, where liquid-liquid and liquid-vapour coexistence end (see panel (b)). A liquid-liquid-vapour triple point is present at temperatures slightly above the temperature of the liquid-vapour critical point. Below $p_c^{(3)}$ the liquid-vapour critical point disappears but the liquid-liquid critical point is present down to $p^{(-)}\approx0.43p_c^{(3)}$. At pressures below $p^{(-)}$ the liquid-liquid demixing becomes metastable with respect to the liquid-vapour transition (see panel (c)) and the phase diagram is similar to those of the $3_A-4_A$ or $3_A-5_A$ mixtures.

\begin{figure}
\epsfig{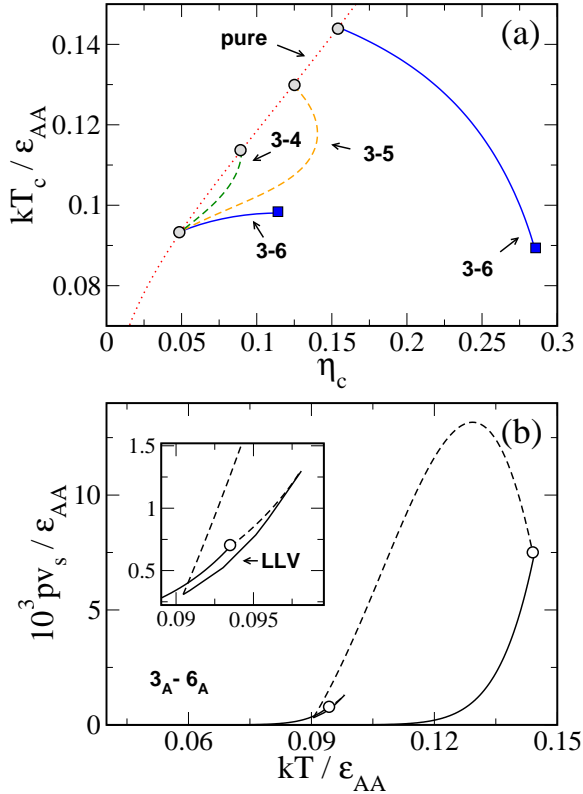}
\caption{(a) Scaled-temperature vs packing fraction at the critical lines of various $3_A-f_A^{(2)}$ mixtures. The critical line of a hypothetical pure fluid with a non-integer number of bonding sites is also represented as a red-dotted line. Grey circles denote the critical points of pure fluids with $3$, $4$, $5$, and $6$ bonding sites. Blue-squares are the critical end points of the $3_A-6_A$ mixture. (b) Pressure-temperature projection of the phase diagram of the $3_A-6_A$ mixture. Solid curves ending in an open circle are the $LV$ lines of the pure fluids. The other solid line (see zoom in the inset) is the $LLV$ three phase line. The critical lines are dashed.} 
\label{fig11}
\end{figure}

The critical properties of mixtures where species $1$ has three bonding sites are shown in Fig. \ref{fig11} (a). For $3_A-4_A$ and $3_A-5_A$ mixtures the critical line is continuous and connects the critical points of both fluids. In the $3_A-6_A$ mixture, however, the critical line is discontinuous (see also the $pT$ projection in panel (b) of the same figure). A liquid-vapour critical line extends from the critical point of the component with a lower number of bonding sites ($3_A$) and terminates at a three phase line at an upper critical end point. Another liquid-vapour critical line starts at the critical point of the other component ($6_A$), changes its character continuously to a liquid-liquid critical line and ends on the same three phase line at a lower critical end point (type V binary mixture). 

\begin{figure}
\vspace{0.1in}
\epsfig{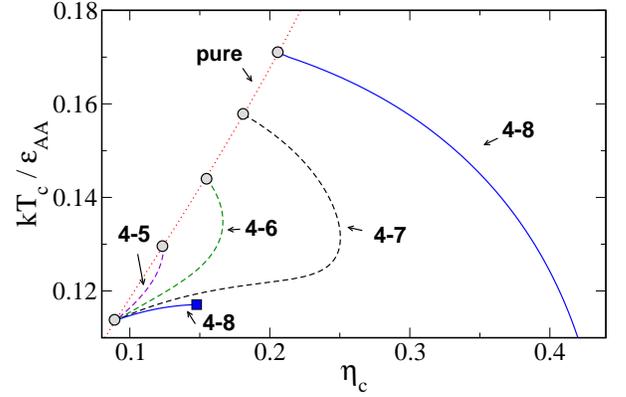}
\caption{Scaled-temperature vs packing fraction at the critical points of $4_A-f_A^{(2)}$ mixtures. The behaviour of a hypothetical pure fluid with a non-integer number of bonding sites is also represented as a red-dotted line. Grey circles denote the critical points of pure fluids with an integer number of bonding sites in the range $4$ to $8$. Critical end points of the $4_A-8_A$ mixture are denoted by blue-squares.} 
\label{fig12}
\end{figure}

The same topological change occurs in mixtures where species $1$ has $4$ bonding sites (see Fig. \ref{fig12}) when species $2$ has $8$ bonding sites. 

We have not studied mixtures with $f^{(1)}_A>4$, as the predictions of Wertheim's theory become less accurate as the number of bonding sites increases \footnote{In addition, in mixtures where both species have the same diameter there is a natural limit to the number of sites per particle that bond simultaneously: the maximum coordination number $12$}, and also because the density is not sufficiently low to rule out the presence of stable solid phases that will appear at high pressure and/or low temperature. However, the stability of the solid is very sensitive to the location of the bonding sites: a symmetric distribution ({\it e.g.}, tetrahedral symmetry in particles with $4$ bonding sites) will promote the formation of crystalline phases, whereas the opposite occurs if the bonding sites are randomly distributed. In Wertheim's theory the bonding sites are independent, and thus it will describe more accurately particles with bonding sites that are distributed randomly. 

\section{\label{Conclusions} Discussion and concluding remarks}

We have analysed the phase behaviour, percolation threshold and critical properties of binary mixtures of patchy colloidal particles, using Wertheim's theory and an extension of the Flory-Stockmayer theory of percolation for binary mixtures. We have restricted the study to mixtures where all bonding sites are identical (there is a single bonding probability) and the particles have the same diameter (no drive for phase separation from the excluded volume of the monomers). Despite the simplicity of the model, we have found a rich phase behaviour including closed miscibility gaps, liquid-liquid phase separation and topological changes in the phase diagram, as the number of bonding sites of the particles is varied. 

We have found that the entropy of bonding plays a crucial role in the stability of fluid phases and drives novel liquid-liquid phase transitions, not present in mixtures of simple fluids. Consistent with this observation the closed miscibility gaps and the liquid-liquid demixing occur deep in the percolated region of the phase diagram. It is then two network or structured fluids that coexist at these new phase transitions. 

The difference in the number of bonding sites of both species was found to be the key parameter that controls the topology of the phase diagram and the nature and number of the fluid phase transitions of the mixture. In particular, we found that:

1. The phase diagram is type I (the critical line of the mixture connects the critical points of the pure fluids) if the species with a higher number of bonding sites has less than twice the the number of bonding sites of the other species. If this condition is not satisfied, then the entropy of bonding drives a new liquid-liquid phase separation and the mapping to a pure fluid with average functionality fails. Furthermore, if both species have finite $LV$ critical points the phase diagram changes from type I to type V (A $LV$ critical line extends from the critical point of the fluid with the smaller number of bonding sites and terminates at a three phase line at an upper critical end point. Another $LV$ critical line starts at the critical point of the other species, changes its character continuously to a liquid-liquid critical line and ends on the same three phase line at a lower critical end point). 

2. The critical line is monotonic if the difference between the number of bonding sites of the two species is one:
\begin{equation}
|f^{(1)}_A-f^{(2)}_A|=1.
\end{equation}
This is the case for $2_A-3_A$, $3_A-4_A$ and $4_A-5_A$ mixtures, and the corresponding fluids are completely miscible at all pressures. The mapping to a pure fluid with average functionality is (qualitatively) correct. By contrast, the critical line is non-monotonic if the difference between the number of bonding sites of the two species is greater than one: 
\begin{equation}
|f^{(1)}_A-f^{(2)}_A|>1.
\end{equation}
Examples are $1_A-3_A$, $2_A-4_A$, $3_A-5_A$ or $4_A-6_A$ mixtures. In this case closed miscibility gaps appear at high pressures and the mapping to a pure fluid with average functionality fails. 

Changes in the topology of the phase diagram of binary mixtures have been reported experimentally and theoretically for a variety of systems. For example, the 
CO$_2+n-$alkane homologous series exhibits transitions from type I to type II, IV and III as the chain length of the hydrocarbon increases \cite{Schneider19985,*Miller1989295}. These changes are generally correlated to the ratio of critical temperatures and pressures of the pure fluids. In the binary mixtures discussed here these ratios also vary when the number of bonding sites varies. We believe, however, that the driving force for the change in this class of mixtures is the difference in the bonding entropy associated with bonds between like- and unlike- particles. It is possible to address this question by considering mixtures of particles with different diameters (or different types of bonding sites) to constrain both fluids to have similar critical pressures and temperatures. 

Related studies on the global phase diagrams of binary mixtures using Wertheim-based equations of state can be found in the literature. See, for example, Ref. \cite{Nezbeda1999193}, where  binary mixtures of water and $n-$alkanols are modeled by the statistical associating fluid theory (for a recent review see, for example, Ref. \cite{reviewSAFT}). In these models, however, the attractive interactions are not limited to the interaction between bonding sites as considered in the present study. This limit is unlikely to be relevant for molecular fluids but appears to be relevant to the newly synthesized patchy particles, and gives rise to new and interesting phenomenology. 

We have already studied the empty fluid regime of binary mixtures of particles with $2$ and $3$ patches of different types \cite{heras:104904} with the goal of designing structured fluids, with novel macroscopic properties, including stable bigel phases \cite{goyal:064511,*B907873H}. 

As a final remark, we note that a state-dependent functionality $\overline{f}$ may be defined by mapping the moments of the cluster distributions functions of the mixture, to those of pure fluids, similarly to what is done in \cite{tavares3}. A pure fluid characterised by a state dependent functionality may be useful to describe aspects of the mixture phase behaviour, such as closed miscibility gaps or liquid-liquid phase separation, that are not described by the mapping to 
a fluid with average functionality. 

\section{Acknowledgments}

This work has been supported, in part, by the Portuguese Foundation for Science and Technology (FCT) through Contracts Nos. POCTI/ISFL/2/618 and PTDC/FIS/098254/2008, by the R$\&$D Programme of Activities (Comunidad de Madrid, Spain) MODELICO-CM/S2009ESP-1691, and by the Spanish Ministry of Education through grant FIS2008-05865-C02-02. D. de las Heras is supported by the Spanish Ministry of Education through contract No. EX2009-0121.

\end{document}